\documentclass[conference]{IEEEtran}
\IEEEoverridecommandlockouts
\usepackage{cite}
\usepackage{amsmath,amssymb,amsfonts}
\usepackage{algorithmic}
\usepackage{graphicx}
\usepackage{textcomp}
\usepackage{xcolor}
\usepackage{comment}
\usepackage{makecell}
\usepackage{color}

\begin{document}

\title{Detection and mitigation of indirect conflicts between xApps in Open Radio Access Networks}

\author{\IEEEauthorblockN{Cezary Adamczyk}
\IEEEauthorblockA{\textit{Poznan University of Technology}\\
Poznań, Poland \\
cezary.adamczyk@doctorate.put.poznan.pl}
\and
\IEEEauthorblockN{Adrian Kliks}
\IEEEauthorblockA{\textit{Poznan University of Technology}\\
Poznań, Poland \\
adrian.kliks@put.poznan.pl}
\thanks{This is the author’s version of an article that has been published in the proceedings of the 2023 IEEE INFOCOM conference.}}

\IEEEpubid{Copyright © 2023 IEEE. Personal use is permitted, but republication/redistribution requires IEEE permission.}

\maketitle

\begin{abstract}
In Open Radio Access Networks, the Conflict Mitigation component, which is part of the Near\mbox{-}RT RIC, aims to detect and resolve any conflicts between xApp decisions. In this paper, we propose a universal method for detecting and resolving of indirect conflicts between xApps. Its efficiency is validated by extensive computer simulations. Our results demonstrate that, in the considered scenario, the mean bitrate satisfaction of users increases by 2\%, while the number of radio link failures and ping-pong handovers in the network is significantly reduced.
\end{abstract}

\begin{IEEEkeywords}
O-RAN, Near-RT RIC, conflict detection, conflict mitigation
\end{IEEEkeywords}

\section{Introduction}
\IEEEPARstart{A}{n} Open Radio Access Network (O\mbox{-}RAN) is a concept of a radio access network (RAN) designed to be more open and interoperable than existing, conventional RANs. It uses open interfaces and distributed architecture to allow different vendors to build components that are compatible with each other, reducing the cost and complexity of deploying and managing RAN base stations (BS).

Near\mbox{-}Real\mbox{-}Time RAN Intelligent Controllers (Near\mbox{-}RT RICs) are key components in O\mbox{-}RAN. These controllers orchestrate the RAN in near-real time, with control loops between 10 ms and 1 s long, and are designed to monitor, predict, and optimize the network's behavior by dynamically adapting to the current conditions \cite{oran_wg1_arch}. Near\mbox{-}RT RICs enable O\mbox{-}RAN to be more agile and adaptive than traditional RANs.

The network optimization features of Near\mbox{-}RT RICs are enabled by various xApp applications \cite{oran_wg1_uc_report}, which are used to support specific network control use cases. xApps can influence specific RAN components (i.e., E2 nodes) using E2 Control messages sent via the E2 interface.

\section{xApp decision conflicts}
Conflicts between xApp decisions occur when multiple E2 Control messages contradict each other. Technical Specifications provided by O\mbox{-}RAN Alliance distinguish three types of these conflicts: direct conflicts (control decisions targeting the change of the same parameter), indirect conflicts (control decisions targeting the change of various, yet related, parameters that affect the same area of network operation), and implicit conflicts (control decisions provided by xApps targeting different optimization goals interfere with each other, having a negative, non-obvious effect on network performance) \cite{oran_wg3_nRT-RIC-arch}.

Each control conflict can negatively impact O-RAN performance and reliability \cite{understanding_oran}. Ideally, conflicts should be completely avoided when a mobile network operator (MNO) chooses which xApps to deploy in the network. However, since xApps can be provided by various third parties, and dependencies between these xApps may not be simple to determine, there is still a need for conflict mitigation measures. The Conflict Mitigation component in the Near\mbox{-}RT RIC is responsible for providing these measures \cite{oran_wg3_nRT-RIC-arch}. Currently, there are no established frameworks and methods to reliably detect and resolve all types of xApp conflicts.

\IEEEpubidadjcol

\section{Indirect conflict detection and resolution}
\begin{figure}[!t]
\centering
\includegraphics[width=0.41\textwidth,angle=0]{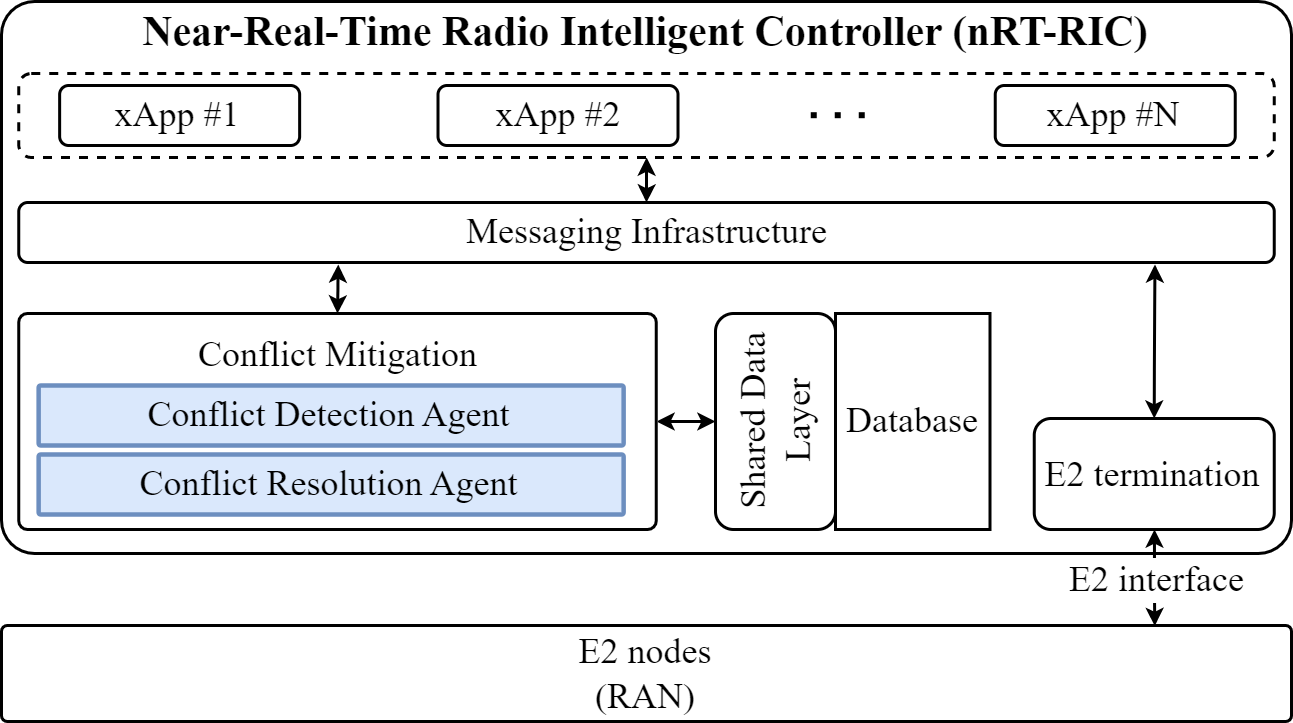}
\caption{Conflict Detection and Resolution framework in the Near\mbox{-}RT RIC}
\label{fig_framework}
\end{figure}

As the original contribution in this article, we propose a conflict detection and resolution scheme for indirect conflicts between xApp decisions. It aims to provide a reliable method of indirect conflict mitigation, applicable to any combination of xApps deployed in the Near\mbox{-}RT RIC. The proposed solution involves splitting the Conflict Mitigation component into two logical parts: Conflict Detection (CD) and Conflict Resolution (CR) Agents, which, as the names suggest, are responsible for conflict detection and resolution, respectively. As a prerequisite for the agents to work as intended, all E2 Control messages from active xApps need to be routed through the Conflict Mitigation component. It is assumed that the Messaging Infrastructure in the Near\mbox{-}RT RIC is configured to meet this requirement. The O\mbox{-}RAN components relevant to the proposed scheme are shown in Figure \ref{fig_framework}.

Indirect conflicts can be detected pre-action by having knowledge about the groups of parameters that have an effect on the same area of RAN operation. For example, xApp \#1 wants to balance cell load by modifying Cell Individual Offset, while xApp \#2 wants to change the cell's electrical tilt. Both control decisions influence the effective cell boundary, causing an indirect conflict. The Conflict Mitigation component needs to be aware of the relations between parameters to detect such conflicts. These relations can be tracked with Parameter Groups (PG), which collect parameters that influence the same area of network operation in relation to a specific control target (i.e., cell, bearer, or user). These groups can be configured manually by the MNO, predetermined in the O\mbox{-}RAN standards, or learned dynamically during network operation.

The CD Agent component is able to detect indirect conflicts between xApp decisions as it monitors all E2 Control messages provided by active xApps. Each incoming E2 Control message is analyzed by the CD Agent, which determines whether any indirect conflicts occur by comparing the incoming message's control target and parameter with the affected parameter groups of the past E2 Control messages that were allowed to impact the network configuration and are still in effect. If the control decision is not blocked to mitigate a conflict, information about the E2 Control message is saved in the Near\mbox{-}RT RIC's database.

When a conflict is detected, a simple conflict resolution method is introduced. The CR Agent is configured to prioritize one of the xApps, and in the case of a conflict, all other xApps' decisions are rejected. The non-prioritized xApps are then put on a cooldown period during which their decisions cannot influence the control target.

\section{Evaluation of the solution}
A heterogeneous O\mbox{-}RAN network is considered with a macro BS (height 28 m, EIRP 42 dBm, frequency 800 MHz, bandwidth 10 MHz) collocated with a micro BS (height 12 m, EIRP 26 dBm, frequency 2100 MHz, bandwidth 20 MHz) in the middle of an urban environment area, and six micro BSs located around it in a hexagonal grid. All BSs are split into three 120 degree-wide sectors. Within each BS, a number of randomly placed users are located -- 100 within the macro BS and 30 within each micro BS. The users move randomly within the considered simulation area. Each user is randomly assigned one of three profiles (low/medium/high throughput, with probabilities 40\%/30\%/30\%, respectively) that defines the required bitrate. The Near\mbox{-}RT RIC is deployed with two xApps working simultaneously: Mobility Load Balancing (MLB) and Mobility Robustness Optimization (MRO). The MLB xApp monitors the percentage load of cells in the network as a ratio of user-assigned PRBs to all PRBs. Based on this percentage, Cell Individual Offset (CIO) is modified to balance the traffic load across nearby BSs (higher load of a cell corresponds to a higher CIO value). Handovers between cells are optimized with the MRO xApp, which adjusts the handover Time-To-Trigger (TTT) and handover hysteresis (HH) of cells to reduce the number of unnecessary handovers, especially "ping-pong handovers". Both xApps indirectly conflict with each other, as all affected parameters (CIO, TTT, HH) influence effective handover boundaries. The simulation was performed for three variants of network operation - with no CM, with CM and prioritization of MLB, with CM and prioritization of MRO. A period of 200 seconds was considered for each variant. Simulation results for all scenarios are shown in Table \ref{tab:sim_results}.

\begin{table}[!t]
\renewcommand{\arraystretch}{1}
\caption{Simulation results}\label{tab:sim_results}
\centering
\begin{tabular}{|l|l|l|l|}
     \cline{2-4}
     \multicolumn{1}{c|}{} & \makecell{MRO + MLB} & \makecell{prio. MLB\\+ MRO} & \makecell{prio. MRO\\+ MLB} \\
     \hline
     \makecell{mean base\\station load} & \textbf{81.23}\% & 83.88\% & 82.71\% \\
     \hline
     \makecell{mean user\\satisfaction} & 63.27\% & 63.02\% & \textbf{65.50\%} \\
     \hline
     \makecell{call blockade\\count} & 547 & \textbf{508} & 542 \\
     \hline
     \makecell{radio link\\failure count} & \textbf{204} & 217 & \textbf{204} \\
     \hline
     \makecell{total handover\\count} & 5630 & 5701 & \textbf{5415} \\
     \hline
     \makecell{ping-pong\\handover count} & 3371 & 3476 & \textbf{3273} \\
     \hline
\end{tabular}
\end{table}

Simulation results show that enabling the proposed conflict mitigation scheme has a number of positive impacts on network operation. When the MRO xApp is prioritized, most network statistics improve compared to CM-less network operation. Mean user satisfaction improves by over 2 p.p., while counts of negative events are significantly decreased (call blockades, handovers, ping-pong handovers) or remain the same (radio link failures). In contrast, prioritizing MLB does not provide as clear of an improvement. On one hand, it significantly decreases the call blockade count, as MLB can freely balance the load between cells. On the other hand, it deteriorates all other metrics to varying degrees -- for example, a slight decrease is observed for mean user satisfaction, while the number of handovers is significantly increased.

\section{Conclusions and next steps}
Evaluation of the proposed conflict mitigation scheme demonstrates potential in improving network performance and reliability without requiring additional resources. Simulations for both considered CM modes show that the benefits of introducing CM measures can drastically vary depending on the scenario. This suggests that any future application would require assessing the exact impact on the network, taking into account the specific MNO's optimization goals.

Next steps in the research should consider other CM approaches (also for direct and implicit conflicts), conflicts between rApps and xApps, and tests for other network scenarios.

\bibliographystyle{ieeetr}
\bibliography{literature}

\end{document}